\documentclass{easychair}
\usepackage[numbers,sort]{natbib}
\usepackage{xurl}
\usepackage{amsmath}
\usepackage{bm}
\usepackage[utf8]{inputenc}
\usepackage{pgfplots}
\usepackage{hyperref}
\usepackage[capitalise]{cleveref}
\pgfplotsset{compat=1.17}

\title{Towards the shortest DRAT proof\\of the Pigeonhole Principle}

\author{
    Isaac Grosof
\and
    Naifeng Zhang
\and
    Marijn J.H. Heule
}

\institute{
  Carnegie Mellon University,
  Pittsburgh, Pennsylvania, United States\\
  \email{\{igrosof,naifengz,marijn\}@cmu.edu}
 }

\authorrunning{I. Grosof, N. Zhang, and M.J.H. Heule}

\titlerunning{Towards the shortest DRAT proof of the Pigeonhole Principle}

\newcommand{\negate}{\overline}
\newcommand{\lra}{\leftrightarrow}

\begin{document}

\maketitle

\begin{abstract}
The Pigeonhole Principle (PHP) has been heavily studied in automated reasoning, both theoretically and in practice.
Most solvers have exponential runtime and proof length, while some specialized techniques achieve polynomial runtime and proof length.
Several decades ago, Cook
manually constructed $O(n^4)$ extended resolution proofs,
where $n$ denotes the number of pigeons.
Existing automated techniques
only surpass Cook's proofs
in similar proof systems
for large $n$.
We construct the shortest known proofs of PHP in the standard proof format of modern SAT solving, DRAT. Using auxiliary variables and by recursively decomposing the original program into smaller sizes, we manually obtain proofs
having length $O(n^3)$ and leading coefficient $5/2$.
\end{abstract}

\section{Introduction}

Many important SAT instances are known to be unsatisfiable,
but are challenging to solve quickly.
A natural way to compare different approaches
is to compare proof length in a given proof system.
This also  allows us to directly compare automated and manual solving efforts.
Finding shorter and shorter proofs of challenging UNSAT instances can serve as a guide for research
into new techniques for efficient SAT solving.
If a short proof exists, we can hope to eventually build a fast solver based on the same principle.

The Pigeonhole Principle (PHP),
when phrased as a SAT instance,
is famously difficult to prove,
making it a good challenge problem.
Haken \cite{HAKEN1985297} proved that
any resolution proof of the Pigeonhole Principle must be exponential in size.
Likewise, state-of-the-art SAT solvers such as {\tt CaDiCaL} and {\tt Kissat} \cite{biere2020solvers}
essentially only search for resolution proofs
for PHP instances, thus requiring exponential time and proof size.

As a result, the Pigeonhole Principle has received extensive research focus over the years
\cite{HAKEN1985297,beame1996simplified,cook1979relative,razborov2001proof,audemard2010restriction,biere2008picosat,biere2014detecting,ignatiev2017tackling}, 
often employed as a way to evaluate stronger proof systems and new SAT solving tools.
It has been known since the 1970s that shorter proofs of unsatisfiability exist, using more powerful proof systems.
Cook \cite{cook1976short} gave an $O(n^4)$ proof of unsatisfiability,
using the extended resolution proof system \cite{tseitin1983complexity}.
His proof consists of $n$ inductively defined PHP formulas,
each of $O(n^3)$ size.
However, the Pigeonhole Principle can be encoded more efficiently, using only $O(n^2)$ clauses.
This gives hope for an $O(n^3)$ size proof along the lines of Cook's proof.
Of course, we must still use the standard inefficient encoding of the input Pigeonhole instance,
only using the more efficient encoding for the inductively defined formulas.

More recently,
a few specialized automated tools have been created which find shorter proofs of unsatisfiability,
empirically scaling as $O(n^3)$,
though with larger leading constants.
Heule, Kiesl, and Biere \cite{heule2017short} describe a novel
extension-free proof system called ``Propagation Redundancy'' (PR),
in which they find $O(n^3)$ length proofs
for Pigeonhole Principle formulas.
Heule and Biere \cite{heule2018difference}
then give a method to convert these
PR proofs into $O(n^3)$ length DRAT proofs.
In a recently published paper,
Bryant, Biere, and Heule \cite{bryant2022clausal}
describe a solver based on Pseudo-Boolean binary decision diagrams (PGBDD),
which outputs DRAT proofs of the Pigeonhole Problem
whose clause length empirically scales as $O(n^3)$.

While these solvers' proofs empirically scale as $O(n^3)$,
the leading constant varies dramatically.
The PGBDD-based solver's proofs have a much larger leading constant than Cook's
$O(n^4)$ clause proof,
so they are only shorter for $n \ge 128$.
The PR-based solver, when its PR proofs converted to DRAT proofs, has a proof length of approximately $3.3n^3$ \cite{heule2018difference}.
This represents the current state-of-the-art in short DRAT proofs for PHP, and it is the benchmark we will attempt to improve upon.

We compare these proofs
by examining their proof lengths using the DRAT proof system \cite{wetzler2014drat},
which has become the standard proof system in SAT solving.
For instance, DRAT has been used as the proof system the SAT Competition
since 2014 \cite{belov2014proceedings}.
Proofs in the DRAT format can be automatically converted
by the tool \texttt{dram-trim} \cite{wetzler2014drat}
into the LRAT proof format \cite{cruz2017efficient},
which can be efficiently checked by several formally verified proof checkers \cite{heule2017efficient}.

We therefore ask:

\begin{quote}
    What is the shortest DRAT proof of unsatisfiability for the standard PHP($n$) formula?
\end{quote}

We give the shortest known DRAT proofs of unsatisfiability for PHP($n$),
both the shortest for concrete small $n$ and for asymptotic $n$.

In this paper, we provide the following contributions:
\begin{itemize}
    \item In \cref{sec:proof}, we give the shortest known DRAT proof of the unsatisfiability of PHP($n$),
    and prove that it is a valid DRAT proof.
    We also give a reference to our implementation of our proof.
    \item In \cref{sec:counting-clauses}, we give an exact formula for the proof length of our proof,
    demonstrating that our proof length scales as $\frac{5}{2} n^3 + O(n^2)$.
    \item In \cref{sec:empirical}, we empirically compare our proof, Cook's original proof,
    and the proofs output by state-of-the-art CDCL-based solvers.
\end{itemize}
\section{Background}
\label{sec:background}

\subsection{CNF Formulas}

We consider propositional formulas in \emph{conjunctive normal form} (CNF), which are defined as follows. 
A \emph{literal} is either a variable $x$ (a \emph{positive literal}) or the negation $\overline x$ of a variable $x$ (a \emph{negative literal}). 
The \emph{complement} $\overline l$ of a literal $l$ is defined as $\overline l = \overline x$ if $l = x$ and as $\overline l = x$ if $l = \overline x$. 
A \emph{clause} is a finite disjunction of the form $(l_1 \lor \dots \lor l_n)$ where $l_1,\dots,l_n$ are literals. 
A \emph{formula} is a finite conjunction of the form $C_1 \land \dots \land C_m$ where $C_1, \dots, C_m$ are clauses. 
For example, $(x \lor \overline y) \land (z) \land (\overline x \lor \overline z)$ is a formula consisting of the clauses $(x \lor \overline y)$, $(z)$, and $(\overline x \lor \overline z)$.
Formulas can be viewed as sets of clauses, and clauses can be viewed as sets of literals.

A \emph{unit clause} is a clause that contains only one literal. 
The result of applying the \emph{unit-clause rule} to a formula $F$ is the removal of all clauses that are satisfied by unit clauses and the removal of all literals that are falsified by unit clauses. 
The iterated application of the unit-clause rule to a formula, until no unit clauses 
are left, is called \emph{unit propagation}. If unit propagation on a formula $F$ yields the empty clause $\bot$, we
say that it derived a \emph{conflict} or a \emph{contradiction} on $F$.
For example, unit propagation derives a conflict on $F =  (\overline x \lor y) \land (\overline y) \land (x)$ since
the unit clauses remove both literals in the first clause, thereby reducing it to $\bot$.

\subsection{Background on DRAT}

We use the DRAT \cite{wetzler2014drat} proof system,
which has become the standard proof system in SAT solving.
In this section, we provide some background
on the DRAT proof system.

The DRAT proof system operates by starting with a CNF formula $F$.
Each line of the proof is either a clause addition instruction, or a clause deletion instruction.
We define a ``working formula'' $F_i$
for every proof line $i$ in the DRAT proof.
The initial formula $F_0$ is simply the input formula $F$.

Given a working formula $F_i$
after $i$ proof lines,
the next working formula $F_{i+1}$
is either $F_i \cup \{ C \}$,
if some clause $C \not\in F_i$ is added,
or $F_i \setminus \{ C' \}$,
if some clause $C' \in F$ is deleted.
A DRAT proof terminates with the addition of the empty clause, demonstrating a contradiction.

A clause $C$ is valid to add or delete
if it satisfies the \emph{Resolution Asymmetric Tautology} (RAT)
property for some literal $l \in C$ with respect to the current formula $F_i$.

A clause $C$ has the RAT property for a literal $l \in C$ with respect to the formula $F_i$
if all resolvents of $C$ on $l$ are implied by $F_i$ via unit propagation.
Specifically, consider all clauses $D \in F_i$ such that $\negate l \in D$.
The resolvent $C \bowtie D$ is defined as:
\begin{align*}
    C \bowtie D := (C \setminus l) \cup (D \setminus \negate l)
\end{align*}
$C$ has the RAT property on literal $l$ if and only if
$F_i \vdash_1 C \bowtie D$ for all such $D$.
That is, $F_i$ implies $C \bowtie D$ via unit propagation.
A formula $F$ implies a clause $C'$ via unit propagation
if the conjunction of $F$ with the negation of each literal $l' \in C'$ leads to a contradiction via unit propagation.

If a clause $C \not \in F$ has the RAT property with respect 
to $F$, then $F$ and $F \cup \{C\}$ are satisfiability equivalent \cite{heule2013verifying}.

As an example of the RAT property, let $F = (a \vee \negate b) \land (\negate a \vee b) \land (b \vee \negate c) \land (c)$.
Let's examine whether the clause $C = (a)$ has the RAT property with respect to $F$, on the literal $a$.
The only clause in $F$ which contains $\negate a$ is $D = (\negate a \vee b)$.
The resolvent is $C \bowtie D = (b)$.
Now, we must check whether $F$ implies $(b)$ via unit propagation.
To do so, we perform unit propagation on $F \cup (\negate b)$.
We find that $\negate c$, followed by a contradiction.
As a result, $F \vdash_1 (b)$, implying that $C$ has the RAT property.

\subsection{Background on the Pigeonhole Principle}

The Pigeonhole Principle states that
\begin{quote}
    It is impossible to put $n+1$ pigeons in $n$ holes,
    with at most one pigeon in each hole.
\end{quote}

To phrase this as Boolean formula,
we will have variables
$x_{ph}$ for each pigeon $p \in [0, n]$
and each hole $h \in [1, n]$,
where $x_{ph}$ represents whether pigeon $p$ is in hole~$h$.
Next, we need to encode the constraints that ``each pigeon is in at least one hole''
and ``each hole contains at most one pigeon''.
The standard CNF encoding of this problem,
given by Cook \cite{cook1976short},
encodes ``each pigeon is in a hole''
as a single clause for each pigeon:
\begin{align*}
 (x_{p1} \lor \dots \lor x_{pn}) \quad \forall  0 \leq p \leq n
\end{align*}
To encode ``each hole contains at most one pigeon'',
the standard CNF encoding
has a clause for each pair of variables corresponding
to the same hole:
\begin{align}
    \label{eq:standard-constraint}
    \negate x_{ph} \vee \negate x_{qh}
    \quad \text{for all pigeons } 0 \leq p < q \leq n, \text{ for all holes } 1 \le h \le n
\end{align}
To denote this standard encoding, we write PHP($n$).
Note that the standard encoding uses $O(n^3)$ clauses.

Cook \cite{cook1976short} gave a proof that the Pigeonhole Principle is unsatisfiable using the Extended Resolution proof system,
using $O(n^4)$ clauses to do so.
Cook's proof proceeds by reducing PHP($n$) to PHP($n-1$).
Cook introduces new variables $x'_{ph}$
for $0 \leq p < n$, $1 \leq h < n$ with definitions as follows:
\begin{align}
    \label{eq:cook-definition}
    x'_{ph} \lra x_{ph} \vee (x_{nh} \wedge x_{pn})
\end{align}
Cook then uses reduction to derive clauses identical to the standard encoding
on the $x'_{ph}$ variables,
encoding an instance with one fewer pigeon and one fewer hole.
Cook then recurses until $n=1$, at which the instance can be immediately
proven unsatisfiable with resolution.
Cook's proof uses $O(n^3)$ clauses in the recursive step from the $n$ to $n-1$ size instances, for a total of $O(n^4)$ clauses.

In this paper and in our linked generator,
we manually construct the shortest known DRAT proofs of the unsatisfiability of the Pigeonhole Problem formula PHP($n$). Our proofs combine the $O(n^3)$ scaling of recent solvers
\cite{heule2017short,bryant2022clausal}
with the immediate generation and small leading constant of Cook's proof.

\section{High-level Overview}

Our goal is to give the shortest known DRAT proof of unsatisfiability 
for the Pigeonhole Principle formula,
where proof length is measured by the number of clauses added.

We want to use the same recursive-variable-introduction style of proof that Cook used,
introducing new formulas of smaller and smaller size.
This is a very effective technique for a short proof, especially for small $n$, such as $n \le 100$.
Among all such proofs in this style, we want to give the shortest possible proof.
Moreover, we want to do so without changing the initial encoding of the pigeonhole formula.

To generate a smaller proof,
we change the encoding of the ``at most one pigeon per hole'' constraint
in the recursively introduced formulas, without changing the initial encoding.
Both the standard PHP($n$) formula and Cook's proof
use the pairwise encoding of this constraint.
There are several more efficient encodings of the at most one constraint,
but we choose the most efficient encoding (the smallest sum of the number of clauses plus the number of variables),
by recursively removing three literals and adding an auxiliary variable.
We describe this more in \cref{sec:detailed}.
As is, these clauses cannot be introduced via DRAT.
To overcome this, we introduce a small number of auxiliary clauses -- only one per auxiliary variable.

Other than the encoding of the ``at most one pigeon per hole'' constraint,
there is little else to improve, with respect to our goal of minimizing the number of clauses added.
The base variables are defined using four clauses each,
which is the minimum for any definition other than a simple AND or OR of two literals.
Such a simple definition
does not seem sufficient for the new formula to have the appropriate structure.
Finally, there are the ``each pigeon is in a hole'' constraints, which are a single clause per pigeon.
Other than the small number of auxiliary clauses, there seems to be little room for improvement.

One possibility for further improvement that we did not explore
would be to remove multiple pigeons in a single step of variable introduction.
It is possible that this could shorten the proof further.
Another possibility is that for very large $n$, another proof style might be shorter,
as one of our style's advantages is its simplicity, which is primarily beneficial at small $n$,
e.g. $6 \le n \le 100$.
We leave these questions to future research.
We believe that our proof is the best possible proof of unsatisfiability for PHP
that uses the recursive-variable-introduction style and removes one pigeon at a time.

\section{Our proof of the unsatisfiability of the Pigeonhole Principle formula}
\label{sec:proof}

In this section,
we specify the clauses we add,
as well as why adding each clause is a valid RAT step.
We describe our proof here in words,
but we have also implemented our proof
in code.
Our proofs are output by the proof generator\footnote{\url{https://github.com/isaacg1/pigeonhole/blob/main/prove-amo-general.py}}.
To generate our proof of PHP($n$),
run \texttt{python3 prove-amo-general.py <n> 3},
where \texttt{<n>} is the desired problem size.
This outputs our proof in the DRAT format.
While our proof generator outputs clause deletions, the deletions are not necessary for our DRAT proof,
and we do not include them in our proof length. They are merely added to speed up verification.

For convenience, we have also provided a generator of the standard encoding
of the Pigeonhole Problem\footnote{\url{https://github.com/isaacg1/pigeonhole/blob/main/generate.py}}.
To generate the problem, run \texttt{python3 generate.py <n>}.
Note that this initial encoding of the Pigeonhole Problem uses the same standard encoding as Cook's proof, making our proof directly comparable to Cook's proof.

To verify our proof, use a tool such as \texttt{drat-trim}\footnote{\url{https://github.com/marijnheule/drat-trim}} \cite{wetzler2014drat}. 

\subsection{Detailed overview of our solution}
\label{sec:detailed}

Recall from \eqref{eq:standard-constraint}
that to encode the constraint that at most one pigeon is in hole $h$,
Cook uses a direct encoding that for each hole $h$, for all $p$, $q$ pigeons, $\negate x_{ph} \vee \negate x_{qh}$, which has a size of $O(n^3)$ clauses.
Let us denote this constraint as $\operatorname{AMO}(x_{0h}, x_{1h}, \ldots, x_{nh})$, where $\operatorname{AMO}$ stands for ``At Most One.'' 

In our encoding, we recursively decompose the AMO constraint
by removing the first three literals, namely $x_{0h}, x_{1h}, x_{2h}$, and add an auxiliary variable $y_{0h}$.
We define $y_{0h}$ to hold if none of the first three literals hold:
\begin{align*}
    y_{0h} \lra \negate x_{0h} \land \negate x_{1h} \land \negate x_{2h}
\end{align*}
We also add pairwise constraints to ensure no pair of the first three literals hold.
Finally, we add $\negate y_{0h}$ to the rest of literals
to form another $\operatorname{AMO}$ constraint
to be recursively decomposed.

Therefore, for each hole $h$, $\operatorname{AMO}(x_{0h}, x_{1h}, \ldots, x_{nh})$ can be decomposed into 
\begin{align}
\nonumber
&\operatorname{AMO}(\negate y_{0h}, x_{3h}, x_{4h}, \ldots, x_{nh}) \\
\label{eq:positive_outline}
&\land (y_{0h} \vee x_{0h} \vee x_{1h} \vee x_{2h}) \\
\nonumber
&\land (\negate y_{0h} \vee \negate x_{0h})
\land (\negate y_{0h} \vee \negate x_{1h})
\land (\negate y_{0h} \vee \negate x_{2h}) \\
\nonumber
&\land (\negate x_{0h} \vee \negate x_{1h})
\land (\negate x_{0h} \vee \negate x_{2h})
\land (\negate x_{1h} \vee \negate x_{2h})
\end{align}

Note that \eqref{eq:positive_outline}
is not necessary to represent the constraint that at most one of $x_{0h}$, $x_{1h}$, $\ldots$, $x_{nh}$
is true.
Instead, by adding the auxiliary clause \eqref{eq:positive_outline},
we constrain the variables $x_{0h}, x_{1h}, x_{2h}, y_{0h}$
to require that exactly one is true.
With this additional constraint
we ensure that unit propagation will function smoothly,
which is key to short DRAT proofs.
If we had an ``at most one'' constraint instead of an ``exactly one''
constraint, the proof would require case analysis,
preventing us from efficiently adding DRAT clauses.

We recursively decompose the latter part,
$AMO(\negate y_{0h}, x_{3h}, x_{4h}, \ldots, x_{nh})$,
following the same procedure.
In the end, our encoding has a size of $O(n^2)$ clauses,
$O(n)$ clauses per hole $h$.
A precise count is given in \cref{sec:counting-clauses}.

To reduce from  PHP($n$) to PHP($n-1$), we remove pigeon $n$ and hole $n$,
introducing new variables $x'_{ph}$
according to the same definition that Cook \cite{cook1976short} used,
given in \eqref{eq:cook-definition}.
In the same fashion as in Cook's proof,
we repeat this reduction $n$ times until the proof is trivial.

Specifically, our proof consists of the following steps:
\begin{itemize}
    \item Iterate by $k$ from $n-1$ down to $1$,
    \begin{itemize}
        \item Definitions:
            \begin{itemize}
                \item Introduce the $x_{ph}$ variables, for  $1 \le h \le k, 0 \le p \le k$.
                \item Introduce the $y_{gh}$ variables, for $1 \le h \le k, 0 \le g \le \lfloor k/2 \rfloor$.
            \end{itemize}
            These clauses are extended resolution clauses, a less-powerful subset of RAT.
        \item Derivations: Prove pairwise constraints between variables in group $g$ other than $y_{gh}$.
        These clauses use the full power of RAT.
        \item Finally,
            for each pigeon $p \in [0, k]$,
            we derive the ``at least one'' constraints
            which specify that each pigeon being in a hole.
            These clauses are RUP clauses, a less-powerful subset of RAT \cite{wetzler2014drat}.
    \end{itemize}
\end{itemize}

Throughout this proof, whenever we write a DRAT clause,
we write the resolution variable first.
We also \textbf{bold} the resolution variable.

In \cref{sec:proof-definition,sec:proof-derived,sec:proof-each}, we describe the process of variable introduction for the general recursive step
where $k < n-1$, in which both the prior formula and the newly introduced formula use our novel encoding.
For the case of $k = n-1$, where the prior encoding is the standard encoding PHP($n$) given in \eqref{eq:standard-constraint},
the same DRAT introduction clauses suffice, though the proof is significantly simpler.

\subsection{Definition Clauses}
\label{sec:proof-definition}
\subsubsection{Introducing Base Variables}

Our first kind of clauses introduce the new $x'_{ph}$ variables.

These clauses are defined in the same fashion as in Cook's proof \eqref{eq:cook-definition}.
There are two ways for pigeon $p$ to be in hole $h$ on iteration $k$:
\begin{itemize}
    \item If pigeon $p$ was in hole $h$ on iteration $k+1$.
    \item If pigeon $p$ was in hole $k+1$ on iteration $k+1$ and pigeon $k+1$ was in hole $h$ on iteration $k+1$.
\end{itemize}
Intuitively, we are deleting pigeon $k+1$ and hole $k+1$,
and moving the pigeon from the deleted hole to the opening created by the deleted pigeon.
Our proof is flexible enough to allow any pigeon to be deleted
-- we delete the $k+1$th pigeon for simplicity of indexing and to mirror Cook's proof.

Symbolically, we define $x'_{ph}$ as follows:
\begin{align*}
    x'_{ph} \lra x_{ph} \vee (x_{(k+1)h} \wedge x_{p(k+1)})
\end{align*}

We implement this definition with the following 4 clauses.
In each case, the redundant variable for DRAT purposes is $x'_{ph}$, the newly introduced variable.
\begin{align}
    \label{eq:neg1}
    &\bm{\negate x'_{ph}} \vee x_{ph} \vee x_{p(k+1)} \\
    \label{eq:neg2}
    &\bm{\negate x'_{ph}} \vee x_{ph} \vee x_{(k+1)h} \\
    \label{eq:pos1}
    &\bm{x'_{ph}} \vee \negate x_{ph} \\
    \label{eq:pos2}
    &\bm{x'_{ph}} \vee \negate x_{p(k+1)} \vee \negate x_{(k+1)h}
\end{align}

One minor optimization that we have made is that
if $p = k$, we omit \eqref{eq:neg1} and \eqref{eq:neg2},
the two clauses above in which
$\negate x'_{ph}$ appears.
Intuitively, these clauses only prevent pigeons from being added from nowhere in iteration $k$, which does not harm the proof.
They are needed for unit propagation steps in RAT proofs when $p < k$,
but because our propagation proceeds towards lower $p$s,
these clauses are unnecessary when $p = k$, the maximum possible
value of $p$.

\subsubsection{Introducing Auxiliary Variables}

Having introduced the $x'_{ph}$ variables,
we now introduce the auxiliary variables $y'_{gh}$.
Each such variable corresponds to a group of $x'_{ph}$ variables.
There are 3 types of groups:
\begin{itemize}
    \item the initial group, group $0$
    \item intermediate groups
    \item and the final group, group $\lfloor n/2 \rfloor - 1$
\end{itemize}

Group $0$ consists of 
\begin{align}
    \label{eq:group-list}
    x'_{0h}, x'_{1h}, x'_{2h}, y'_{0h}
\end{align}
Intermediate groups $g \in \{1, 2, \ldots,  \lfloor n/2 \rfloor - 2\}$
consist of 
\begin{align*}
\negate y'_{(g-1)h}, x'_{(2g+1)h}, x'_{(2g+2)h}, y'_{gh}
\end{align*}
The final group consists of $\negate y'_{(\lfloor n/2 \rfloor - 2)h}$ 
and the remaining $x'$ variables.
If $n \le 3$, there is only one group, consisting only of $x'$ variables, which we think of as a final group.

For each group which is not a final group,
we have 7 clauses.
Of these, 4 are involved in introducing $y'_{gh}$.
Let $l_1, l_2, l_3$ be the three literals in the group.
Symbolically, we define $y'_{gh}$ as
\begin{align*}
    y'_{gh} \lra \negate l_1 \land \negate l_2 \land \negate l_3
\end{align*}
To introduce $y'_{gh}$, we add the clauses
\begin{align}
    \label{eq:positive}
    &\bm{y'_{gh}} \vee l_1 \vee l_2 \vee l_3 \\
    \nonumber
    &\bm{\negate y'_{gh}} \vee \negate l_1 \\
    \nonumber
    &\bm{\negate y'_{gh}} \vee \negate l_2 \\
    \nonumber
    &\bm{\negate y'_{gh}} \vee \negate l_3
\end{align}
Note that \eqref{eq:positive} is not logically required to encode the ``at most one pigeon per hole'' constraint,
but it is necessary to prevent a case analysis that would disrupt the RAT clause addition.

\subsection{Derived clauses}
\label{sec:proof-derived}

In this section, we handle the derived clauses within each group, which are not simply definitions.
We describe this for the general case where $k < n-1$.
When $k = n-1$,
we add the same DRAT clauses,
but the prior encoding is the standard encoding,
so the proof that the added DRAT clauses are valid is slightly different,
but similar and simpler.

The three remaining clauses relating to group $g$ are:
\begin{align}
    \label{eq:first-hard}
    &\bm{\negate x'_{ph}} \vee y'_{(g-1)h} \\
    \label{eq:second-hard}
    &\bm{\negate x'_{(p+1)h}} \vee y'_{(g-1)h} \\
    \label{eq:third-hard}
    &\bm{\negate x'_{(p+1)h}} \vee \negate x'_{ph}
\end{align}
where $p = 2g+1$.
Note that for the initial group $g=0$,
$y'_{(g-1)h}$ is replaced by $x'_{0h}$.

Intuitively, these clauses are valid to add because if we assume their negations,
we can perform unit propagation through variables in iteration $k+1$
to conclude that no other pigeons were present in iteration $k+1$,
then move to iteration $k$ and unit propagate back up to pigeon $p$,
eventually reaching a contradiction.

To see why these hold in more detail, let us look at their resolvents.
Let us consider the first clause, involving $x'_{ph}$, first.
The only clauses where $x'_{ph}$ has appeared positively are in the introduction of $x'_{ph}$, specifically \eqref{eq:pos1} and \eqref{eq:pos2},
and the four-literal clause from the introduction  of $y'_{gh}$,
\eqref{eq:positive}.

Let us restate these clauses:
\begin{align}
    \nonumber
    &x'_{ph} \vee \negate x_{ph} \\
    \nonumber
    &x'_{ph} \vee \negate x_{p(k+1)} \vee \negate x_{(k+1)h} \\
    \label{eq:tautology}
    &y'_{gh} \vee \negate y'_{(g-1)h}  \vee x'_{ph} \vee x'_{(p+1)h}
\end{align}
The resolvent of \eqref{eq:tautology}
and \eqref{eq:first-hard} contains $y'_{(g-1)h}$ and $\negate y'_{(g-1)h}$,
so it is an immediate tautology.

We therefore must show that the remaining two clauses, after resolving with \eqref{eq:first-hard}, are implied by unit propagation:
\begin{align}
    \label{eq:resolve1}
    &y'_{(g-1)h} \vee \negate x_{ph} \\
    \label{eq:resolve2}
    &y'_{(g-1)h} \vee \negate x_{p(k+1)} \vee \negate x_{(k+1)h}
\end{align}

For example, let's look at \eqref{eq:resolve1}.
To prove it via unit propagation, we want to show that assuming
$\negate y'_{(g-1)h} \wedge x_{ph}$
yields a contradiction via unit propagation.

Let us focus on the assumption that $x_{ph}$ holds.
Using unit propagation on the clauses introduced in iteration $k+1$, we can conclude that
$\negate x_{q h}$ for all $q \neq p$.

Next, using the clauses that introduced the definitions of $x'_{q h}$
in iteration $k$, we can conclude that $\negate x'_{q h}$ for all $q < k$.

Now, we can conclude that $y'_{0h}$,
using the four-literal clause \eqref{eq:positive} from group 0.
We can then derive $y'_{1h}$ from the four-literal clause
\eqref{eq:positive} from group 1,
and so on, deriving that $y'_{f h}$ for all $f < g$.

In particular, we conclude that that $y'_{(g-1)h}$,
using the four-literal clause \eqref{eq:positive} from group $g-1$.
This produces the desired contradiction.

We can derive a similar contradiction to prove the second desired clause.
From the assumption that $x_{p(k+1)}$,
we use unit propagation to prove that $\negate x_{q (k+1)}$ for all $q \neq p$.
From the assumption that $x_{(k+1)h}$,
we use unit propagation to prove that
$\negate x_{q h}$ for all $q < k+1$.

Now, we can once again conclude that
\begin{align*}
    &\negate x'_{q h} \quad \forall q < p\\
    &y'_{f h} \quad \forall f < g
\end{align*}
We thereby reach the same contradiction.

We have now given a RAT proof that it is valid to add \eqref{eq:first-hard}
at this point in the proof.
The RAT proofs for the other two clauses, \eqref{eq:second-hard} and \eqref{eq:third-hard}, are essentially identical.

\subsection{Deriving ``each pigeon is in a hole'' clauses}
\label{sec:proof-each}

Finally, for each pigeon $p$,
we add the clause 
\begin{align*}
    x'_{p1} \vee x'_{p2} \vee \ldots \vee x'_{pk}.
\end{align*}

This clause $C$ satisfies the RUP condition, which means that the current formula $F$
implies $C$ itself via unit propagation, without needing to look at any resolvents.
To see why,
assume that $\negate x'_{ph}$ for all $h \in [1, k]$.
We previously introduced clauses of the form
\begin{align*}
    x'_{ph} \vee \negate x_{ph}
\end{align*}
From unit propagation, we find that $\negate x_{ph}$
for all $h \in [1, k]$.

From iteration $k+1$, we have a clause which says that
\begin{align*}
    x_{p1} \vee \ldots x_{pk} \vee x_{p(k+1)}
\end{align*}
As a result, $x_{p(k+1)}$ must hold.
But we also previously introduced clauses of the form
\begin{align*}
    x'_{ph} \vee \negate x_{p(k+1)} \vee \negate x_{(k+1)h}
\end{align*}

We can therefore conclude that $\negate x_{(k+1)h}$ must hold, for all $h \in [1, k]$.
We therefore conclude that $x_{(k+1)(k+1)}$, by parallel reasoning.
Now, we have concluded that both $x_{p(k+1)}$ and $x_{(k+1)(k+1)}$.
This is two pigeons in the same hole, so the rest is straightforward.
Specifically, we can conclude that the $y_{g(k+1)}$ variable
for the group $g$ containing pigeon $p$ must hold,
and hence none of the $y_{g'(k+1)}$ for $g' > g$ can hold
eventually reaching a contradiction on the final group,
which contains pigeon $k+1$.
As a result, this clause is valid to add.

\section{Counting Clauses}
\label{sec:counting-clauses}

In this section, we count the exact number of DRAT clauses
used in our proof, as well as Cook's proof \cite{cook1976short}
as implemented into DRAT by Randy Bryant's generator\footnote{\url{https://github.com/rebryant/pgbdd/blob/master/benchmarks/pigeon-cook.py}}.

\subsection{Our proof}

For a given iteration $k$,
three types of clauses are added.

For each $p \in [0, k], h \in [1, k]$,
we have a newly defined variable $x_{ph}$.
To define the variable,
four clauses are added.
However, for defining the final pigeon $p=k$,
we only use two clauses.
There are a total of $(4k+2)k$
such clauses.

Next, we have the group clauses,
which encode the ``at most one pigeon per hole'' constraint.
There are $\lfloor \frac{k}{2} \rfloor$
groups
for the $k+1$ pigeons in a given hole.
The first and last groups have up to 3 ``$x$'' variables,
while all intermediate groups have 2.
For each group that is not the final group,
we have 7 clauses.
For the final group,
we have $3$ clauses if
$k \equiv 1 \pmod 2$,
and $6$ clauses if $k \equiv 0 \pmod 2$.
Let $f(k)$
be the number of group clauses per hole,
which has value:
\begin{align*}
    f(k) = \left\lfloor \frac{7}{2} k \right\rfloor - 4
\end{align*}
As an exception, $f(1) = 1$.
In each case, the number of group clauses is $kf(k)$.

Finally, we have an ``at least one clause'' for each pigeon.
There are $k+1$ such clauses.

In total, the number of clauses added in iteration $k$,
for $k>1$ is
\begin{align*}
    k(4k+2) + k\left( \left\lfloor \frac{7}{2} k \right\rfloor - 4 \right) + k+1
\end{align*}
Such iterations are performed for $k$ from $n-1$
down to $2$,
with 9 clauses for $k=1$
and 1 empty clause to complete
the proof.

We can therefore
calculate the exact number
of clauses used by our proof,
for all $n>1$:
\begin{align*}
    &\frac{5}{2}n^3 - \frac{35}{8}n^2 + \frac{11}{4}n + 2
    \text{ if } n \equiv 0 \pmod 2 \\
    &\frac{5}{2}n^3 - \frac{35}{8}n^2 + 3n
    + \frac{15}{8}
    \text{ if } n \equiv 1 \pmod 2
\end{align*}

\subsection{Cook's proof}

For a given iteration $k$,
three types of clauses are added.

For each $p \in [0, k], h \in [1, k]$,
we have a newly defined variable $x_{ph}$.
To define the variable,
four clauses are added.
There are a total of $4(k+1)k$
such clauses.

For each pair of new variables
$x_{ph}, x_{qh}$,
such that $0 \le p < q \le k,
1 \le h \le k$,
2 clauses are added.
There are $(k+1)k^2$ such clauses.

Finally, for each pigeon $p \in [0, k]$,
there is an ``at least one'' constraint.
There are $k+1$ such clauses.

In total, the number of clauses added
in iteration $k$ is
\begin{align*}
    4(k+1)k + (k+1)k^2 + k+1
    = k^3 + 5 k^2 + 5k + 1
\end{align*}
Such iterations are performed for each $k$
from $n-1$ down to $1$,
and one final empty clause is added to complete the proof.

The exact number of clauses used by Cook's proof is:
\begin{align*}
    \sum_{k=0}^{n-1} k^3 + 5 k^2 + 5k + 1
    = \frac{1}{4}n^4 + \frac{7}{6}n^3
    + \frac{1}{4}n^2 - \frac{2}{3}n
\end{align*}

\section{Empirical Results}
\label{sec:empirical}

\begin{figure}
\centering
\begin{tikzpicture}
\begin{axis}[
    xlabel={$n$},
    ylabel={Number of clauses},
    xmin=1, xmax=100,
    ymin=1,
    ymode=log,
    xmode=log,
    legend pos=south east,
    ymajorgrids=true,
    grid style=dashed,
    width=12cm,
    height=8cm,
    smooth,
    thick,
    legend cell align=left,
]

\addplot [ 
    color=cyan!80!black,
    mark=square*
    ] 
    table [
    x=n, 
    y=cook, 
    col sep=comma,
    each nth point=3,
    ] {php.csv};
    \addlegendentry{Cook's Proof}

\addplot [ 
    color=orange!80!white,
    mark=triangle*
    ] 
    table [
    x=n, 
    y=amo3, 
    col sep=comma,
    each nth point=3
    ] {php.csv};
    \addlegendentry{Our Proof}


\addplot [ 
    color=green!80!black,
    mark=diamond*,
    ] 
    table [
    x=n, 
    y=cadical-direct, 
    col sep=comma
    ] {php.csv};
    \addlegendentry{Direct, \texttt{Cadical}}

\addplot [ 
    color=green!50!black,
    mark=diamond*,
    ] 
    table [
    x=n, 
    y=cadical-amo, 
    col sep=comma
    ] {php.csv};
    \addlegendentry{AMO, \texttt{Cadical}}

\addplot [ 
    color=red!80!white,
    mark=*,
    ] 
    table [
    x=n, 
    y=kissat-direct, 
    col sep=comma
    ] {php.csv};
    \addlegendentry{Direct, \texttt{Kissat}}

\addplot [ 
    color=red!50!white,
    mark=*,
    ] 
    table [
    x=n, 
    y=kissat-amo, 
    col sep=comma
    ] {php.csv};
    \addlegendentry{AMO, \texttt{Kissat}}
    
\end{axis}
\end{tikzpicture}
\caption{Number of clauses in proofs of pigeonhole problem for n holes.}
\label{fig:php-comparison}
\end{figure}
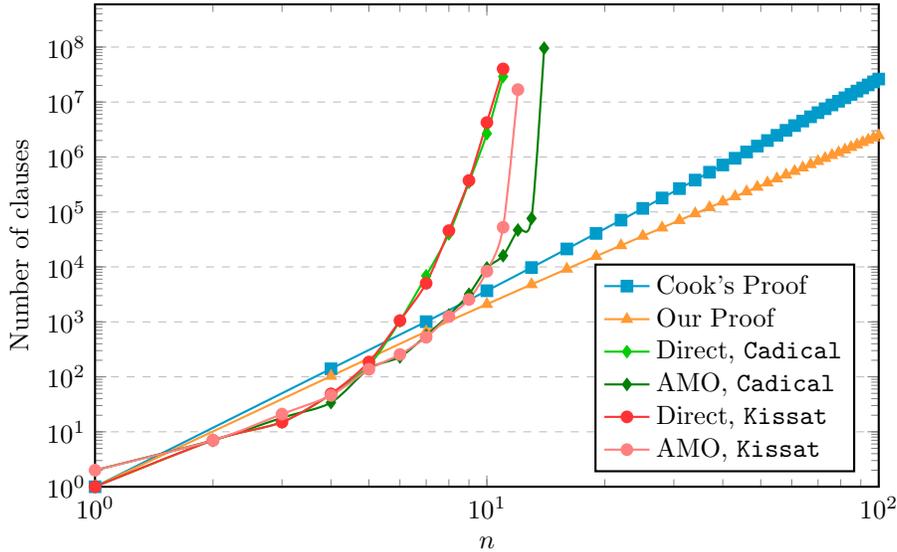

We verified our proof using the \texttt{drat-trim} tool \cite{wetzler2014drat}
for $n$ up to $71$, at which point each verification took more than 10 minutes.

We compared the length of our proof against Cook's proof and two state-of-the-art CDCL-based SAT solvers, \texttt{CaDiCaL}\footnote{\url{https://github.com/arminbiere/cadical}} and \texttt{Kissat}\footnote{\url{https://github.com/arminbiere/kissat}}.
To each solver, we input both Cook's direct encoding (the standard encoding)
as well as our recursive AMO encoding.
We generated Cook's proof using Randy Bryant's generator\footnote{\url{https://github.com/rebryant/pgbdd/blob/master/benchmarks/pigeon-cook.py}}. 
We measured the number of clauses generated by each approach for proofs from PHP(1) to PHP(100) for Cook's proof and our proof.
For the solvers, we ran all $n$
for which the solver completed in at most 10 minutes.

The number of clauses in the proofs output by
\texttt{CaDiCaL} and \texttt{Kissat}
grow exponentially with the instance size $n$.
This can be seen in Fig~\ref{fig:php-comparison},
as the curves are growing superlinearly on the log-log plot.
While the proofs are shorter when the AMO encoding is given as input,
they still grow in length exponentially,
and become far longer than our proof and Cook's proof
once $n$ is greater than 10.

Both our proof and Cook's proof form straight lines in Fig~\ref{fig:php-comparison},
showing that the number of clauses grows polynomially with $n$.
The clause counts exactly match our formulas from \cref{sec:counting-clauses}.

For all $n > 1$, our proof is shorter than Cook's proof.
In the asymptotic limit,
the ratio between Cook's proof length and our proof length
converges to $n/10$.
This follows from the fact that our leading term is $\frac{5}{2}n^3$,
while Cook's leading term is $\frac{1}{4}n^4$.
For example, for $n=100$,
our proof adds $2,456,527$ clauses,
while Cook's proof adds $26,169,100$ clauses.
Our proof is 10.65 times shorter than Cook's proof for $n=100$.

\section{Optimization for a small number of pigeons}

In \cref{fig:php-comparison} in \cref{sec:empirical},
we showed that our proof is the shortest known proof
for $n$ larger than $7$.
For inputs $n \le 7$,
using {\tt CaDiCaL} directly, combined with {\tt drat-trim}'s \texttt{-O} optimization flag,
gives the shortest known proof.

This allows an opportunity to shave a few lines off our proof for $n \ge 7$.
We use our manual proof to reduce from PHP($n$)
to our novel encoding for $k=n-1$, $k=n-2$, and so on. However, instead of recursing all the way to $k=1$,
we stop early after $k=8$, and switch over to a hardcoded proof.

We have implemented this optimization using the flag \texttt{-{}-optimized}
in our proof generator, for inputs $n > 7$.
Using this flag saves 229 steps from our proof,
coming even closer to the shortest proof possible.

\section{Conclusion}

We give the shortest known DRAT proof of the Pigeonhole Principle formula PHP($n$).
Our proof size scales as $O(n^3)$,
with a leading constant of $\frac{5}{2}$,
smaller than any prior proof.
Our proof is asymptotically shorter
than Cook's proof \cite{cook1976short},
the best previously known manually constructed proof valid for all $n$.
Specifically, our proof is shorter by a factor of $\frac{n}{10}$ for large $n$.

While our proof is the shortest DRAT proof known,
there is still room for improvement,
giving rise to several open problems.
Can one achieve a smaller leading asymptotic term than $\frac{5}{2}n^3$?
How much smaller can a proof be for specific $n$?

Another important direction is exploring other proof systems or cost models.
Our proof makes use of the full power of the RAT proof system.
If the proof was restricted to Extended Resolution,
could one adapt our proof to still give a $O(n^3)$ proof?
Many of our proof steps have $O(n)$ length unit propagation sequences.
What if the proof system was changed from DRAT to LRAT,
and every clause hint counted as part of the proof size?
Could an $O(n^3)$ proof still be achieved?

\bibliography{ref} 

\end{document}